\documentclass[%
 aip,
 amsmath,amssymb,
 reprint,%
]{revtex4-1}

\usepackage{graphicx}
\usepackage{dcolumn}
\usepackage{bm}

\usepackage[utf8]{inputenc}
\usepackage[T1]{fontenc}
\usepackage{mathptmx}
\usepackage{etoolbox}
\usepackage{xcolor}
\usepackage{hyperref}

\makeatletter
\def\@email#1#2{%
 \endgroup
 \patchcmd{\titleblock@produce}
  {\frontmatter@RRAPformat}
  {\frontmatter@RRAPformat{\produce@RRAP{*#1\href{mailto:#2}{#2}}}\frontmatter@RRAPformat}
  {}{}
}%
\makeatother

\begin{document}

\preprint{APS/123-QED}

\title{Density-Based Long-Range Electrostatic Descriptors\\for Machine Learning Force Fields}

\author{Carolin Faller} 
\affiliation{University of Vienna, Faculty of Physics \& Vienna Doctoral School in Physics, Kolingasse 14-16, A-1090 Vienna, Austria}
\author{Merzuk Kaltak}
\affiliation{VASP Software GmbH, Berggasse 21, A-1090 Vienna, Austria}
\author{Georg Kresse} 
\email{georg.kresse@univie.ac.at}
\affiliation{University of Vienna, Faculty of Physics, Kolingasse 14-16, A-1090 Vienna, Austria}
\affiliation{VASP Software GmbH, Berggasse 21, A-1090 Vienna, Austria}




\date{%
\today}

\begin{abstract}
This study presents a long-range descriptor for machine learning force fields (MLFFs) that maintains translational and rotational symmetry, similar to short-range descriptors while being able to incorporate long-range electrostatic interactions. The proposed descriptor is based on an atomic density representation and is structurally similar to classical short-range atom-centered descriptors, making it straightforward to integrate into machine learning schemes. The effectiveness of our model is demonstrated through comparative analysis with the long-distance equivariant (LODE) \citep{LODE} descriptor. In a toy model with purely electrostatic interactions, our model achieves errors below 0.1\%, worse than LODE but still very good.\\
For real materials, we perform tests for liquid NaCl, rock salt NaCl, and solid zirconia. For NaCl, the present descriptors improve on short-range density descriptors, reducing errors by a factor of two to three and coming close to message-passing networks.
However, for solid zirconia, no improvements are observed with the present approach, while message-passing networks reduce the error by almost a factor of two to three. Possible shortcomings of the present model are briefly discussed.
\end{abstract}

\maketitle


\section{Introduction}

Machine learning force fields (MLFFs) are a powerful tool for accurately and efficiently predicting inter-atomic potentials, approaching the precision of ab-initio calculations \citep{manzhos2006random, Behler07, Bartok10, rupp2012fast, behler2015constructing, Thompson15, morawietz2016van, bartok2018, SchNet, Csanyi19, ACE, lilienfeld_20, lilienfeld_burke20, mlff_2021, keith_21, NequIP, MACE, ko23, GNoME, universalMACE}. 
MLFFs describe the potential energy as a function of descriptors that represent the atomic structure of a given material.\\
To increase the efficiency of MLFFs, many methods rely on local descriptors that solely represent the atomic environment in the vicinity of an atom within a specified cutoff sphere. These methods assume that all interactions between atoms further apart than the cutoff radius are negligible.  While this approach accurately describes many materials and their properties, 
it neglects long-range interactions such as electrostatics. This is because, in practice, it is impossible to increase the cutoff radius arbitrarily. However, these long-range interactions can be important \cite{lr_import_1, lr_import_2, lr_import_3, lr_import_4}, and new methods that include long-range interactions are necessary to improve the predictive power of MLFFs for systems with long-range electrostatics \cite{4th_gen, spooky_net, tkatchenko_16, tkatchenko_19, tkatchenko_22}.\\

This issue is widely recognized, and various methods have been developed to address it. Recent models have concentrated on specific types of long-range interactions, such as electrostatics, and have introduced correction terms to account for them \cite{LODE, lode2, 4th_gen, spooky_net, car_lr, lilienfeld_18, tensormol, physnet, david_21, gao_22}. A common approach is to split the energy into distinct parts and model them independently. Then, all contributions are added together to calculate the total energy. Short-range contributions are typically calculated using a local model, while long-range components are modeled with diverse approaches. One can either treat the various energy terms separately and train two machine learning (ML) models, adding up the final energies, or utilize combined descriptors and train a single model.\\ 
Many of these methods use classical (non-equivariant and or even non-message passing) 
neural networks (NNs)\cite{4th_gen, car_lr, tensormol, physnet, david_21, gao_22}, or kernel methods \cite{LODE, lode2, lilienfeld_18, huguenin2023physics}. Deep NNs offer more flexibility but require more data for training and are generally slow to train. 
Kernel methods are more convenient for small to medium-sized problems, as they can rely on dense linear algebra routines to solve the linear least squares problem \cite{BishopChristopherM2010Pram}.\\

Another approach to include long-range interactions is to use short-range descriptors, but effectively propagate interactions beyond the distinct cutoff by employing Message Passing Neural Networks (MPNNs) \cite{MPNN1, SchNet, MPNN2, spooky_net, NequIP, MACE, newtonnet, DimeNet, TeaNet, M3GNet, chgnet, GNoME, universalMACE}. Other relevant developments are equivariant MPNNs such as NewtonNet \cite{newtonnet}, DimeNet \cite{DimeNet}, TeaNet \cite{TeaNet}, NequIP \cite{NequIP} and MACE \cite{MACE} that cover not only invariant but also tensorial quantities.\\
NewtonNet is designed to respect Newtonian mechanics, ensuring the learned forces obey physical laws. It also incorporates directional information from forces, enhancing the model's accuracy \cite{newtonnet}. DimeNet employs directional message passing, which allows it to capture complex interactions effectively \cite{DimeNet}. TeaNet uses tensor-based message passing, facilitating the learning of intricate relationships between atomic structures \cite{TeaNet}.\\
NequIP\cite{NequIP} employs equivariant convolutions of tensorial quantities and multi-level message passing of equivariant features, resulting in a particularly 
flexible network topology with many millions of parameters. MACE \cite{MACE} is in most respects a simplified version of NequIP 
usually relying on only two message-passing layers and largely linear activation functions. This should improve execution speed
and learning efficiency. \\

The present work utilizes kernel methods. Our objective is to find a physically meaningful and flexible approach for describing long-range interactions without resorting to a global and non-atom-centered description. Similarly to short-range models \cite{Bartok10,Bartok13,Jinnouchi19a, Jinnouchi19b, desc, jinnouchi2020fly}, the atomic density around each atom is expanded into a suitable set of basis functions. However, we implicitly account for all periodic images of all atoms in the supercell by performing the calculations in reciprocal space and using suitably slowly decaying basis functions.\\
This approach offers the advantage of a flexible, physics-based descriptor that is atom-centered but also long-ranged. It can be easily combined with a local descriptor, as both have the same mathematical form, albeit in our present implementation the approach is not yet very performant.\\
We apply the present approach to a gas of point charges. The descriptor is then compared to the long-distance equivariant (LODE) framework \cite{LODE} and the MPNN MACE \cite{MACE} for liquid sodium chloride and zirconia.

\section{Theory}
\subsection{Short-Range Descriptors}
The descriptors used in this work are designed to resemble the density-based Smooth Overlap of Atomic Positions (SOAP) \cite{Bartok10} and Gaussian Approximation Potential (GAP) \cite{Bartok13} descriptors. The atomic density $\rho$ is calculated around each atom $j$ as a function of the atomic positions denoted as $\textbf{r}$. 
The expansion coefficients
\begin{equation}\label{eq:c}
    c_{nlm}^{jJ} = \sum_{k = 1}^{N_J} h_{nl}\left(r_{jk}\right) Y_{lm}^*\left(\hat{\textbf{r}}_{jk}\right)
\end{equation}
with 
\begin{align}
    h_{nl}\left( r\right) &= \frac{4 \pi}{\left(\sqrt{2\sigma^2\pi}\right)^3} f_{\text{cut}} \left( r\right) \int_0^\infty \chi_{nl} \left( r'\right) \\
    &\times \exp\left( - \frac{r'^2 + r^2}{2 \sigma^2}\right) \iota_l\left( \frac{rr'}{\sigma^2} \right) r'^2 \text{d}r'
\end{align}
from Ref. \onlinecite{Jinnouchi19a} that are given there in Eq. (18) and (19) are used for building the short-range descriptors. Here, the vector $\textbf{r}_{jk}$ is the vector pointing from atom $j$ to $k$, $\textbf{r}_{k}-\textbf{r}_{j}$.
The set of radial $\chi_{nl}$ and angular $Y_{lm}$ basis functions are used to expand the atomic density in order to obtain rotational and translational invariant expansion coefficients \cite{Bartok10,Bartok13}. The indices $l$, $m$, and $n$
are the angular, momentum, and radial index, respectively. The parameter 
$\sigma$ is used to broaden the density distribution. $\iota_l$ are the modified spherical Bessel functions of the first kind, and $N_J$ is
the number of atoms of type $J$. The cutoff function $f_{\text{cut}}$ is used to ensure a smooth decay of $h_{nl}$ to zero at the cutoff radius.\\
 
The final descriptor $\textbf{X}_j$ describing the local environment surrounding atom $j$ is derived by combining the invariant ($l=0$) two-body expansion coefficients 
\begin{equation} \label{eq:cn}
    c^{jJ}_n=c^{jJ}_{n00}
\end{equation}
and the rotational and translational invariant three-body expansion coefficients
\begin{equation} \label{eq:p}
    p^{jJJ'}_{nn'l} = 
    \sqrt{\frac{8\pi^2}{2l+1}}\sum_{m=-l}^{l}c^{jJ}_{nlm}c^{jJ'}_{n'lm}
\end{equation} 
into vectors
\begin{align}
    \textbf{X}_j^{(2)}&=\left(c^{i1}_1, c^{i1}_2, \hdots,  c^{i2}_1, c^{i2}_2, \hdots \right)^T\\
    \textbf{X}_j^{(3)}&=\left( p^{i11}_{110}, p^{i11}_{111}, \hdots,  p^{i11}_{120},  p^{i11}_{121}, \hdots, p^{i12}_{110},  \hdots,  p^{i22}_{110},  \hdots \right)^T 
\end{align}
that are then combined to 
\begin{equation}\label{eq:supervec}
\textbf{X}_j= 
\begin{pmatrix}
\beta^{(2)}\textbf{X}_j^{(2)}\\ 
\beta^{(3)}\textbf{X}_j^{(3)} 
\end{pmatrix} 
\end{equation}
using the weights $\beta$ for the two- and three-body descriptors. In the present calculations $\beta^{(2)} = \sqrt{0.1}$ and $\beta^{(3)} = \sqrt{0.9}$ were used for the short-range descriptors. Whenever inproducts are formed in the kernel, the weights are implicitly squared, so this results in a metric with "weights" of 0.1 and 0.9 for two-body and three-body terms.

\subsection{Long-Range Descriptors}\label{lrdesc}
The objective is to calculate a similar set of descriptors where the expansion coefficients however represent the atomic density over long ranges and multiple supercells. 
The same ansatz as in real space is used, where the expansion coefficients are designed to represent the atomic density around each atom. As before, the descriptors are obtained by projecting the density onto a product basis of radial and angular functions. However, to avoid introducing a cutoff, the corresponding products are evaluated in reciprocal instead of real space:
\begin{equation}\label{eq:clr}
    c_{nlm}^{jJ} =  \sum_{k = 1}^{N_J} \sum_{\textbf{G}} \exp\left( i \textbf{G}\left(\textbf{r}_k - \textbf{r}_j \right) \right) \exp\left( - \frac{|\textbf{G}|^2 \sigma^2}{2}\right) f_{nlm}^*(\textbf{G})
\end{equation}
where \textbf{G} are momentum vectors consistent with the periodic boundary conditions, and $f_{nlm}(\textbf{G})$ are the basis functions in reciprocal space and the other terms represent the broadened atomic density distribution of atom $k$ of type $J$ with positions $\textbf{r}_k$ around atom $j$ with positions $\textbf{r}_j$. The central atom $j$ may be of any atom type present in the material. The translational invariance is obvious from Eq. (\ref{eq:clr}), as seen from the difference vector $\textbf{r}_k - \textbf{r}_j$, where $\textbf{r}_j$ is the position of the central atom.\\
Starting out from the product of the radial $\chi_{nl}\left(r\right)$ and angular $Y_{lm}\left(\hat{\textbf{r}}\right)$ basis functions in real space
\begin{equation} \label{eq:fnlmr}
     f_{nlm}(\textbf{r}) = \chi_{nl}\left(r\right) Y_{lm}\left(\hat{\textbf{r}}\right)
\end{equation}
taking the Fourier Transform (FT) of Eq. \eqref{eq:fnlmr} and applying the plane wave expansion yields 
\begin{equation}
\label{equ:ft}
    f_{nlm}(\textbf{G}) = 4 \pi i^l Y_{lm}(-\hat{\textbf{G}}) \int_0^\infty r^2 j_l(\textbf{G} r) \chi_{nl}(r)  \text{d}r
\end{equation}
where the remaining integral can be either solved analytically or numerically depending on the choice of radial basis functions $\chi_{nl}(r)$.\\

The long-range expansion coefficients given in Eq. \eqref{eq:clr} have the same form as the short-range expansion coefficients given in Eq.  \eqref{eq:c}. Like conventional two- (and three-) body descriptors, they are also atom-centered and invariant under translations and rotations.
So for all practical purposes, they behave and can be used like conventional short-range descriptors.
For example, the learned regression coefficients are transferable to different unit cells and do not depend on the choice of the Bravais lattice.
The only difference is that the evaluation of the expansion coefficients $c_{nlm}^{jJ}$ is performed in reciprocal space. This allows to cover arbitrary interaction ranges, and thus, to encode and measure long-range density fluctuations.\\
With the appropriate choice of basis functions (see next section), the long-range expansion coefficients can encode the long-range density correlations. This is due to the calculation in reciprocal space, where not only the position of an atom is considered but also all periodic images of this atom. It is important to note that these long-range descriptors do not require a cutoff, but of course, it is possible to impose a cutoff by using the FTs of finite-ranged real-space functions in Eq. \eqref{equ:ft}.\\
The individual long-range expansion coefficients are then combined into a vector 
("long-range descriptor"). In the present work, we mostly
 use two-body expansion coefficients as shown in Eq. \eqref{eq:cn} and the corresponding vector $\textbf{X}_j^{(2)}$. The three-body long-range coefficients are excluded since the objective here is to describe long-range pairwise interactions. Furthermore, we found no improvements when including long-range three-body terms in any of the systems considered here. This might have many reasons, but we believe that three-body interactions are generally shorter-ranged. Consider, for instance, the case of Van der Waals (vdW) interactions: two-body vdW interactions fall off like $1/r^6$, and the three-body terms (Axilrod–Teller) fall off like $1/r^9$. 

\subsection{Long-Range Radial Basis Functions}\label{basisfunc}
The type of basis functions used for the descriptors has not been specified yet. For the short-range method, spherical Bessel functions are used for the radial part and spherical harmonics for the angular part.\\ 
Spherical harmonics are also used for the angular part of the long-range descriptor. However, spherical Bessel functions cannot be employed to model the infinitely ranged interactions in the radial part, as to maintain a fixed spatial resolution, the number of Bessel functions would need to increase linearly with $R_\text{cut}$.\\
Given the objective of representing electrostatic interactions that decay slowly like $1 / r$, where $r$ is the distance between two atoms, it is evident that exponentially decaying functions are a sensible choice. This is because 
\begin{equation}\label{eq:slater_1}
\begin{split}
    \frac{1}{r} &
    = \int_0^{\infty} \exp(-\zeta r) \text{d}\zeta \\
    &\approx \sum_{n = 1}^{N_{\text{max}}} w_{n} \exp(-\zeta_n r)
\end{split}
\end{equation}
holds. The first line of Eq. \eqref{eq:slater_1} is the Laplace transform of $1$ resulting in $1 / r$, while the second line approximates this integral using a quadrature rule. This demonstrates that the integral over all exponentially decaying functions can represent $1 / r$, a fact that has been amply used in quantum chemistry to deal with integrals in many-body perturbation theory \cite{LT1, LT2, LT3, LT4, kaltak2014low}. 
It is important to understand that the $1/r$-function can be represented very accurately by the sum of exponentials \cite{hackbusch2019computation}. If both, the decay constants and the weights are optimized independently 6, 8 and 10 functions are sufficient to represent $1/r$ with a maximum absolute error of $10^{-5}$, $10^{-6}$ and $10^{-7}$, respectively, on an interval between $r=[r_\text{min}, 100 \cdot r_\text{min}]$. More functions are required as the required range increases, but the number of functions increases slowly. It is also found that the optimal decay constants are typically exponentially spaced, with the typical factor between successive decay constants being close to two. We, therefore, expect that if the shortest nearest neighbor distance is around, say, 2 \AA, then about 6-8 functions should be sufficient to describe long-range interactions up to typically 200 \AA $ $ with very high accuracy.\\
For simplicity, we here define $\zeta_n$ on a logarithmic mesh where
\begin{equation}
    \zeta_n = \zeta_{\text{max}} / s^{n-1},
\end{equation}
and $\zeta_{\text{max}}$ and the scaling constant $s$ are hyperparameters determined numerically. As already noted, this choice was inspired by our experience with many-body applications where a roughly exponential scaling was also found to be optimal \cite{kaltak2014low}.

\subsection{The LODE Implementation} \label{ourLODE}
Grisafi and Ceriotti introduced the long-distance equivariant (LODE) framework, which uses descriptors to encode the electrostatic potential around atoms \cite{LODE}.  The potential is calculated by summing the potentials induced by other atoms at the position of the central atom.\\
In reciprocal space, the electrostatic potential $\Phi$ is given as the negative product of the atomic density $\rho$ and the Coulomb Kernel 
\begin{equation} \label{eq:coulomb}
    \Phi(\textbf{G}) = - \frac{4 \pi}{\| \textbf{G} \|^2 \Omega} \rho(\textbf{G}).
\end{equation}
where $\Omega$ is the volume of the supercell.\\
Because of the relation shown in Eq. \eqref{eq:coulomb}, it is relatively easy,  with the density-based model presented above, to implement the LODE descriptors. The only required step is to multiply the density-based expansion coefficients by the negative Coulomb kernel before summing over the reciprocal grid points $\textbf{G}$. Spherical Bessel functions are used as radial basis functions for this LODE implementation.\\
As in the original work of Grisafi and Ceriotti, for real materials, the LODE descriptors are combined with short-range descriptors as well, using the approach shown in the following section.

\subsection{Combination of Short- and Long-Range Descriptors}
To apply the models to realistic materials in which both short-range and long-range interactions are present, the final descriptor vectors $\textbf{X}$ must contain information about both types of interaction. By simply appending the short-range descriptors $\textbf{X}_{\text{sr}}$ and long-range descriptors $\textbf{X}_{lr}$ the final descriptor
\begin{equation} \label{eq:comb_descs}
    \textbf{X} = \left[ \textbf{X}_{\text{sr}}^T, \textbf{X}_{\text{lr}}^T \right]^T
\end{equation}
is obtained. This descriptor contains all the necessary information and still provides a unique similarity measure for kernel-ridge regression.\\
As the long-range descriptors are very similar in construction to the short-range descriptors, additional scaling was found to be irrelevant for them. However, to obtain an accurate MLFF involving LODE, we found it crucial to optimize the weight of the LODE descriptors using a hyper-parameter search, as the LODE descriptors have a potentially very different scale as other descriptors. 

\subsection{The Fitting Process}
The fitting process of our model employs the methods of polipy4vasp, the code used in  Ref. \onlinecite{bernhard}. These methods are implemented analogously to those of the Vienna \textit{Ab initio} Simulation Package (VASP) \cite{vasp1, vasp2, vasp3} which are presented in Refs. \onlinecite{Jinnouchi19a, Jinnouchi19b, desc}.\\
The following paragraphs provide a brief overview of the main points of the scheme. It is assumed that there is a non-linear functional mapping of the final descriptors obtained as described above to the total energy of the system. As usual, it is furthermore assumed that the energy $U$ of the system  can be decomposed into the sum 
\begin{equation}
    U = \sum_j U_j
\end{equation} 
of local energy contributions $U_j$ that depend on the "local" environment around each atom $j$.  Kernel regression is used to describe the non-linear dependence of the energy on the descriptors, and a sparse representation is used for the kernel. Either a Gaussian (radial basis function) kernel
\begin{equation}\label{eq:kernel_gauß} 
K(\textbf{X}_j,\textbf{X}_b) = \text{exp}\left(- \frac{\| \textbf{X}_j - \textbf{X}_b \| ^2} {\eta^2} \right)
\end{equation}
or a polynomial kernel
\begin{equation}\label{eq:kernel_poly} 
K(\textbf{X}_j,\textbf{X}_b) = \left(\hat{\textbf{X}}_j\cdot\hat{\textbf{X}}_b\right)^{\zeta}
\end{equation}
is used here.\\
In Eq. \eqref{eq:kernel_poly} $\hat{\textbf{X}}$ indicates that the descriptor $\textbf{X}$ is normalized. The index $b$ in both Eq. \eqref{eq:kernel_gauß} and Eq. \eqref{eq:kernel_poly} refers to the kernel basis function to which the descriptor of the current central atom $\textbf{X}_j$ is "compared".\\

The energy per atom of structure $s$ containing $N_{\text{atom}}$  atoms is obtained when the following equation is fulfilled in the least squares sense:
\begin{equation}\label{eq:U}
    \frac{U^s}{N_{\text{atom}}^s} \stackrel{!}{=} \sum_{j = 1}^ {N_{\text{atom}}^s} \frac{U_j}{N_{\text{atom}}^s} = \sum_{b = 1}^{N_{\text{b}}} w_{b} \sum_{j = 1}^ {N_{\text{atom}}^s} \frac{K(\textbf{X}_j^s,\textbf{X}_b)}{N_{\text{atom}}^s}.
\end{equation}
Here $U^s$ is the energy of structure $s$ obtained from first principles (FP) calculations and $N_b$ is the total number of kernel basis functions. The energy $U^s$ is a linear function of the fitting weights $w_b$. As the forces are the negative derivatives of the energy with respect to the atomic positions, also the forces are linear functions of the weights $w_b$.\\
These linear relations can be expressed as a system of linear equations in matrix-vector form:
\begin{equation}\label{eq:linsys}
    \textbf{y} \stackrel{!}{=} \boldsymbol{\Phi} \textbf{w}.
\end{equation}
The vector $\textbf{y}$ contains the energies of all training structures $s$ and all the forces acting on all atoms in the systems included in the training data set obtained from FP calculations. These entries are made dimensionless by dividing them by the standard deviation of those energies and the forces, respectively.\\
During training the linear system of equations is calculated via the pseudoinverse of $\boldsymbol{\Phi}$ using a singular value decomposition, and singular values smaller than a threshold of $10^{-9}$ times the largest singular value are disposed of. \\ 

\subsection{The Root-Mean Square Percentage Error}

In this work, all errors of training and test data are given in terms of the root-mean-square percentage error (RMSPE). This is calculated by dividing the root-mean-square error of the predicted properties by the standard deviation of the exact results multiplied by $100$. With this approach, we obtain a unitless measure that does not depend on the number of atoms or the size of the unit cell. Typically, errors are given in (m)eV/atom for the energies and in eV/\AA $ $ for the forces. However, these errors cannot be easily compared between different materials or at different temperatures. The RMSPE simplifies such comparisons.\\

\section{Results}
For LODE the model with only pairwise descriptors and a body order term of $\nu = 1$ is used. This facilitates a comparison with the purely radial description of the density-based long-range descriptors. Furthermore, it was found that the use of three-body terms in our case leads to larger errors compared to the use of only two-body terms. 
This is likely related to the already discussed two-body nature of long-range Coulomb interactions.\\

\subsection{Long-Range Effects of Point Charges}\label{randgas}
As an initial test set for the present long-range model, a gas of randomly distributed point charges, analogous to the one used in Ref. \onlinecite{LODE}, was constructed. The set comprises systems of varying volumes, each containing 64 atoms, with 32 having a positive charge of $+1$ and the other 32 having a negative charge of $-1$. To prevent large energies the minimum distance between two atoms was set to $2.5$ \AA. 
The training and validation data were generated using the Ewald energies and forces (as implemented in VASP).
Only long-range descriptors were used for this system.\\
For all tests on this toy model the Gaussian kernel given in Eq. \eqref{eq:kernel_gauß} with a broadening of $\eta = 1.55$ was used.\\

\begin{figure}[h] 
    \includegraphics[width=8cm]{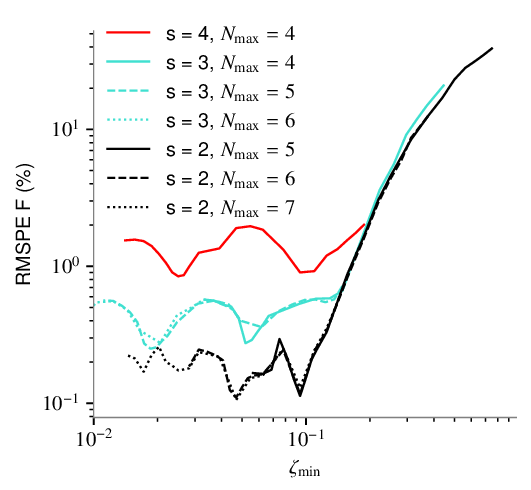}
	\caption{The RMSPE in the forces for various values of the minimal $\zeta_{\text{min}} = \zeta_{\text{max}} / s^{N_{\text{max}} - 1}$ of the radial basis functions, scaling constants $s$, and number of radial basis functions $N_{\text{max}}$ for the gas of point charges.  It is evident that the error reduction is primarily associated with a decreasing scaling constant. All curves with a small scaling constant of $s = 2$ yield a minimum at $\zeta_{\text{min}} \approx 0.09$. Exponents larger than this minimum value are insufficiently long-ranged to capture long-range electrostatic interactions accurately.} \label{fig:zeta}
\end{figure}

The first step is to determine the radial basis functions.  This involves optimizing three parameters: the number of radial basis functions $N_{\text{max}}$, the maximal exponent $\zeta_{\text{max}}$, and the scaling constant $s$. A grid search of these parameters was performed.\\
Figure \ref{fig:zeta} shows the minimal exponents $\zeta_{\text{min}} = \zeta_{\text{max}} / s^{N_{\text{max}} - 1}$ on the x-axis and the corresponding error in the force predictions on the y-axis. Each line represents a distinct set of maximal exponents,  $\zeta_{\text{max}}$, ranging from $12$ to $0.9$ and is associated with a specific value of  $N_{\text{max}}$ and $s$. The linestyle and color of the lines indicate the value of  $N_{\text{max}}$ and $s$, respectively.\\

Overall, reducing the scaling constant $s$ reduces the RMSPE. This is because a smaller scaling constant results in a denser distribution of radial basis functions, allowing for better a description of the electrostatic kernel ($1 / r$). For the electrostatic interactions to be accurately described, it is necessary to have a sufficient number of basis functions such that they decay sufficiently slowly. We see a more or less pronounced minimum around $\zeta_{\text{min}} \approx 0.09$. The second minimum to the left corresponds to the case that the second smallest $\zeta$ becomes approximately $0.09$.
For $\zeta$, the inclusion of values smaller than $0.09$ is unnecessary for this specific training set but also does not significantly degrade the quality of the fit. \\
The plot shows a steep increase in error on the right-hand side, indicating that the smallest exponent is insufficient to capture relevant long-range electrostatic interactions.  This occurs when the radial basis function with the smallest exponent decays too quickly and the basis functions are unable to model the relevant interactions.\\

We found that with a scaling constant of $s = 2$ and 6 radial basis functions, relative errors of approximately $0.1$\% are achieved.\\ 
The values $N_{\text{max}} = 6$, $s = 2$ and $\zeta_{\text{max}} = 2$ are the optimal choices among all possible values for this specific test system, but depend somewhat on the cell size and shortest distance. These hyperparameters will be used in the subsequent calculations. Similar tests must be conducted to determine the optimal set of these three hyperparameters for other materials. However, the values obtained here provide a good starting point, which can help avoid extensive grid searches.\\

Next, we compare our approach with the LODE method, which is implemented in the way presented in \ref{ourLODE}. It is known from theoretical considerations that the two-body LODE descriptor is ideal for a system where only Coulomb interactions are present \cite{LODE}.  Consequently, using only two-body descriptors and a single radial basis function for a small cutoff is sufficient.  In this case one descriptor for each interaction pair (+1, +1), (+1, -1), (-1, +1), and (-1, -1) suffices, as these exactly represent the interaction of one positive or negative point charge with all other positive and negative point charges. In our case, the radial basis function, onto which the potential is projected, is a spherical Bessel function that is smoothly cut off at the cutoff radius.\\
The density-based descriptors are not optimal for this case, so we need to use more radial basis functions and expansion coefficients to learn the Coulomb interactions. However, learning only pairwise interactions is sufficient even using density-based descriptors.\\

\begin{figure*}[t!]
    \includegraphics[width=14cm]{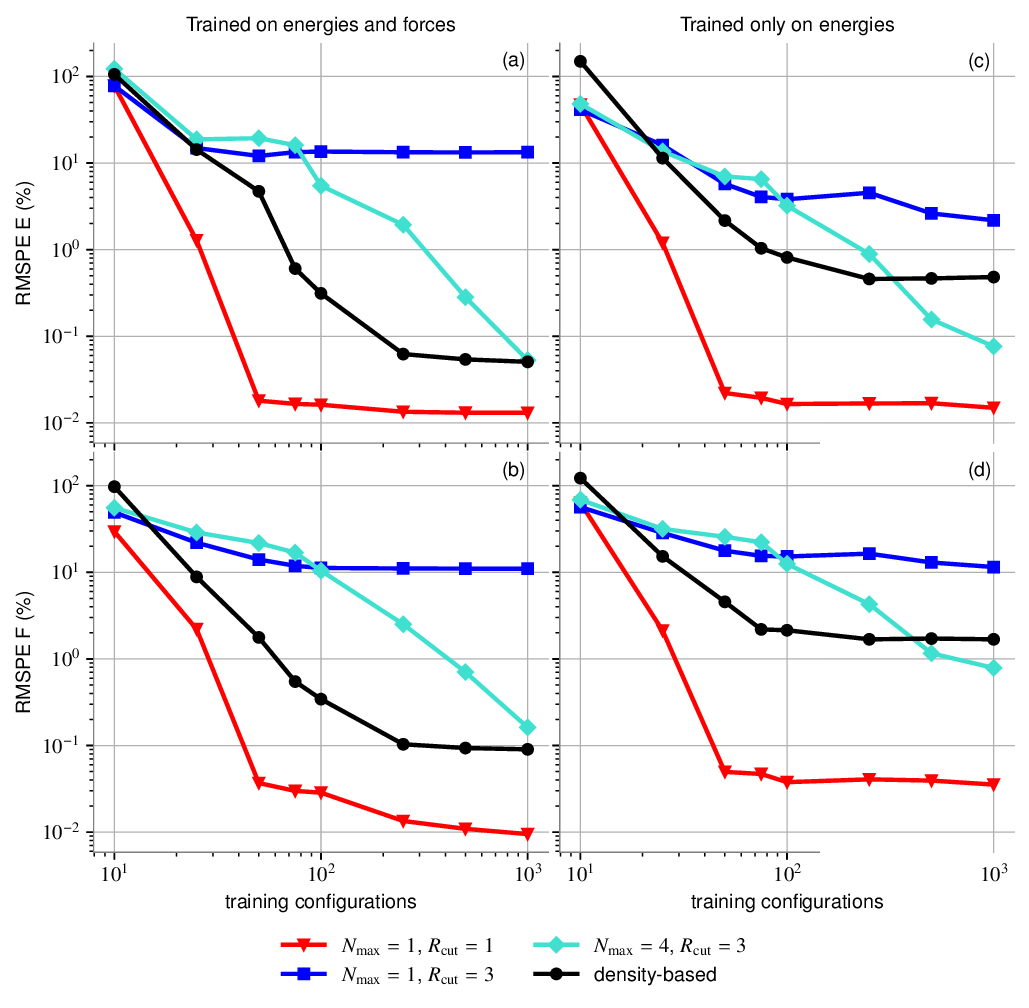}
	\caption[Learning curve for random gas.]{The learning curves for various LODE models using different cutoff-radii $R_{\text{cut}}$ and number of radial basis functions $N_{\text{max}}$ and our density model differ when trained solely on energies. The top panels display the relative percentage error for energy, while the bottom panels present the relative error for the forces. The left panels show the results of training on energies and forces, while the right panels show the results of training on energies only. 
 } \label{fig:rand_gas_lc}
\end{figure*}

Figure \ref{fig:rand_gas_lc} shows the learning curves for the two different methods. The LODE method was tested with various combinations of cutoff radii  $R_{\text{cut}}$ and number of radial basis functions $N_{\text{max}}$, resulting in different numbers of expansion coefficients, and potentially probing the electrostatic potentials at points further away from the central atom. For the LODE model, this adds irrelevant information for this specific simple $1/r$ point charge model.\\ 

The red line represents the ideal LODE descriptor, which quantifies the electrostatic potential inside the sphere of small radius $R_{\text{cut}}$ around the central atom. The electrostatic interaction between two point charges $i$ and $j$ is given by
\begin{equation}
    V_{ij} = \frac{1}{2} \frac{Z_iZ_j}{\| \textbf{r}_i - \textbf{r}_j \|}
\end{equation}
where $\textbf{r}_i$ and $\textbf{r}_j$ are the positions of the particles and $Z_i$ and $Z_j$ are their charges. Since the descriptors include information about each atom type $J$ the charges are not relevant for the implementation and can be set to one. Then the regression will essentially determine $Z_i Z_j$.\\
If one takes now the sum over all pairs $(i, j)$ and over all atoms the electrostatic energy $E$ of the system 
\begin{equation}
    E = \frac{1}{2} \sum_i Z_i \sum_{j \neq i} \frac{Z_j}{\| \textbf{r}_i - \textbf{r}_j \|} = \frac{1}{2} \sum_i Z_i \phi(\textbf{r}_i)
\end{equation}
is obtained as a function of the electrostatic potential at $\textbf{r}_i$
\begin{equation}
    \phi(\textbf{r}_i) = \sum_{j \neq i} \frac{Z_j}{\| \textbf{r}_i - \textbf{r}_j \|}.
\end{equation}
This potential is exactly "measured" by the LODE descriptor with one radial basis function if the cutoff radius is smaller than the minimal atomic distance (the Gaussian broadening adds additional width to the source charges though).\cite{LODE} The use of a cutoff sphere is possible due to Gauss's law: the electrostatic potential at the center of the sphere can be determined by a probe charge of finite size, as long as there is no source term inside the sphere of the probe charge.\\
With a cutoff radius of 3 \AA, the value measured by the single descriptor does not correspond exactly to the value of the potential at the center of the sphere. Hence, the electrostatic potential of point charges that enter the projection sphere can not be determined accurately. This is why the blue line levels off at relatively high RMSPEs. Generally, considering a Gaussian broadening of $\sigma = 0.3$ for the atomic source charges, a cutoff radius of $R_{\text{cut}}  \lesssim 2.5~$\AA$-  5 \sigma$ must be chosen, where $2.5~$\AA\ is the shortest nearest neighbor distance in our models and the factor 5 before $\sigma$ assures that the broadened charge has decayed to negligible values.\\
When using four radial basis functions (turquoise line), a linear combination of the values of the descriptors can approximate a $\delta$-function at the origin, although the model requires additional training data to learn the precise linear combination that corresponds best to the $\delta$-function. \\

When training only on energies and predicting forces (as shown in Fig. \ref{fig:rand_gas_lc} right panels (c) and (d)), the energy predictions are mostly more accurate. However, the errors in the predicted forces are notably higher and seem to stagnate at some point. The same precision as in the mixed training on energies and forces is never reached. This indicates that the models require information about the forces to make highly accurate predictions for forces possible. \\

A comparison of our density-based descriptor (black line) with the ideal LODE descriptor for point charges reveals that the LODE descriptor (red line) is superior. While our density-based descriptor yields satisfactory results, it is necessary to employ more radial basis functions (specifically, six instead of one) and perform a hyper-parameter search to identify the optimal choice of $\zeta_{\text{max}}$, scaling constant $s$, and the number of radial basis functions. This shows that our density-based descriptor, as predicted by theory, is not the best possible descriptor for this prototypical system. Nevertheless, it is still capable of accurately describing the interactions, albeit requiring more training data.\\

The LODE descriptor with one radial basis function and a cutoff radius smaller than the minimal atomic distance (red line) and the density-based descriptor (black line) can both achieve errors below 1\% when sufficient training data are used. The LODE descriptor requires only about 40 training structures to achieve a relative error of 1\%. The density-based descriptor requires 250 training structures to achieve the same accuracy and then stagnates. We believe the stagnation at small errors is likely related to the condition number of the design problem becoming very large, which makes it difficult for the pseudo-inversion to separate "noise" from relevant data.
Nevertheless, both models give very satisfactory results. These are better than those typically obtained using machine learning, as the errors for MLFFs are typically in the mid-single digit percentage range (around 2 to 10\%). The hyperparameters, such as the number of radial basis functions and reciprocal lattice points, can be adjusted to increase the speed and memory efficiency of the computations while still achieving excellent results.

\subsection{Liquid Sodium Chloride}\label{NaCl}

After demonstrating the models' ability to describe systems with purely $1/r$ interactions, they were applied to real materials. Liquid sodium chloride (NaCl) was chosen as the first material due to the significant difference in electronegativity between the Na and Cl atoms suggesting the presence of non-negligible long-range electrostatic interactions. The combined descriptors capture both short- and long-range interactions, enabling analysis of material properties depending on both types of interactions.\\
The dataset for this material consists of 1014 different structures, each containing 64 Na and 64 Cl atoms. The configurations are taken from VASP molecular dynamics (MD) simulations with 50000 steps and a time step of 1.5 fs. During the MD, the material was heated from 1100K to 1400K using a Langevin thermostat. An energy cutoff of 350 eV was used. The PAW potentials used were "PAW\_PBE Na\_pv 19Sep2006" for Na and "PAW\_PBE Cl 06Sep2006" for Cl, considering seven valence electrons for each element.\\

Table \ref{tab:res_nacl} presents the results for NaCl obtained using five different methods: purely local/short-range descriptors (Local), MACE without MP (MACE no MP), LODE combined with short-range descriptors (LODE), long-range density-based descriptors combined with short-range descriptors (Density) and invariant MACE with one MP layer (MACE inv.) and equivariant MACE with one MP layer (MACE equiv.). The combination of short- and long-range descriptors follows the procedure outlined in Eq. \eqref{eq:comb_descs}.\\
For the short-range part of the descriptor, the hyperparameters were set to $\sigma = 0.5$, $R_{\text{cut}}=~$6 \AA, $N_{\text{max}} = 6$ and $L_{\text{max}} = 3$ where $L_{\text{max}}$ corresponds to the angular quantum number and indicates the number of angular basis functions. For the density-based part, the optimal choice was $N_{\text{max}} = 7$, $s = 2$, $\zeta_{\text{max}} = 1.5$ and $\sigma = 0.3$.\\
For the LODE minimal, we used a model that can describe $1/r$ interactions exactly ($R_\text{cut} = $ 1 \AA, $N_{\text{max}} = 1$ and $\sigma = 0.3$) with the long-range part being weighted 40 times more than the short-range part and the forces being weighted 3 times stronger than the energies. A more flexible LODE model was also used, with a larger cut-off radius of $R_\text{cut} = $7 \AA  and $N_{\text{max}} = 5$, where the long-range part was weighted by a factor of 0.003 compared to the short-range part (the LODE descriptors show huge fluctuations in the case of a liquid).\\
A polynomial kernel of order $\zeta = 4$ was used.\\
For MACE $R_{\text{cut}} =~$6~\AA, descriptors with up to four-body interactions, force weights of $1000$, energy weights of $10$ and $100$ epochs were used. After $80$ epochs the energy weights are increased to further lower the error in the energies.\\
Out of the 1014 training configurations always $20\%$ were used as validation data.\\

\begin{table}[h]
\caption{\label{tab:res_nacl}%
The root-mean-square percentage error (RMSPE) was calculated for the validation data of liquid NaCl using combined density and radial LODE descriptors describing the electrostatic potential and a more flexible one as well as local-only descriptors and MACE \cite{MACE} with and without Message Passing (MP) as well as using invariant (inv.) and equivariant (equiv.) features. Radial cutoffs of typically 6~\AA $ $ are used. The results show that the density-based approach outperforms the LODE method for real materials and is as effective as MACE with invariant MP. Nevertheless, including long-range interactions is essential as the purely local scheme and MACE without MP yield large errors.\\}

\begin{ruledtabular}
\begin{tabular}{lcc}
 &RMSPE E (\%)& RMSPE F (\%)\\
 \hline
 Local&7.7&12.6\\
 MACE no MP&8.5&8.5\\
 LODE minimal&5.6 &3.7\\
 LODE flexible&7.0 &11.0 \\
 Density&2.2&3.3\\
 MACE inv.&3.0&3.1\\
 MACE equiv.&2.1&2.2\\
 \end{tabular}
\end{ruledtabular}
\end{table}

The density-based descriptors and Message Passing MACE outperform the purely local method. LODE performs better than the local methods but not as good as the density-based and MP models.\\
The likely explanation for the better performance of the density-based approach is that the LODE is optimal for purely electrostatic interactions, yet lacks the capability to describe intermediate range interactions.  In real materials, the $Z_iZ_j/\| \textbf{r}_{ij} \| = Z_iZ_j/r_{ij}$ Coulomb interaction between particles $i$ and $j$ is screened by the electrons, resulting in an effective interaction 
\begin{equation}
    V = \int \frac{Z_iZ_j}{\left\| \textbf{r}_i - \textbf{r}'\right\|} \epsilon^{-1} \left( \textbf{r}', \textbf{r}_j\right) \text{d}^3 r' .
\end{equation}
At very large distances the interaction will be screened by the ion-clamped macroscopic dielectric constant $\epsilon^{-1}$. However, at medium distance, the electronic screening is less pronounced and the interaction approaches $1/r$. LODE is designed to model a functional form of $1/r$ only but can, by construction, not model screened interactions that deviate from the functional $1/r$ behavior. The present approach is more flexible as the functional form of the interaction is not keyed into the model but rather learned.\\
Even increasing the cutoff radius of the LODE method to 7 \AA $ $ and using more radial basis functions (LODE flexible) does not improve the results. On the contrary, the errors in the forces increase by a factor 1.3,  indicating that the electrostatic potential at points further away from the atom of interest does not add any relevant information to the representation of the atomic environment. \\

\begin{figure}[h]
    \includegraphics[width=8cm]{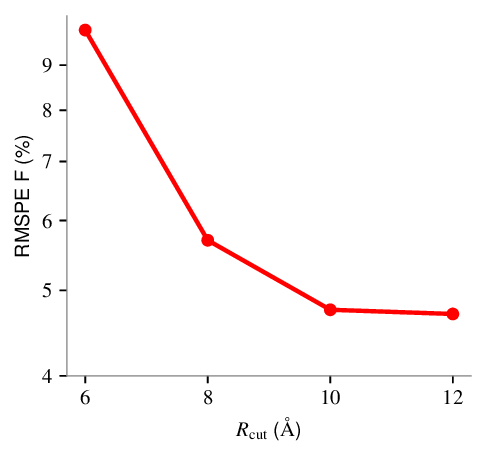}
    \caption{RMSPE in the forces as a function of the two-body cutoff radius $R_{\text{cut}}$ for NaCl for the short-range descriptors. Data was calculated using VASP. The local method cannot match the accuracy of the combined density method or MACE even when the cut-off radius is increased to 12~\AA.}\label{fig:rcut} 
\end{figure}

Even when the cutoff radius $R_{\text{cut}}$ of the purely local method is systematically increased to make the descriptors longer-ranged the result never becomes quite as good as for the combined density method or MACE. This is shown in Fig. \ref{fig:rcut}. These data were calculated using the MLFF implemented in VASP. For the three-body descriptors a cutoff radius of $5$ \AA$ $ was used, since this yields better results than 6~\AA. This is also the reason why the initial value in this plot is better than the one shown in 
Tab. \ref{tab:res_nacl} obtained using polipy4vasp. In polipy4vasp a distinction between two-body and three-body cutoffs is impossible and the best compromise is found to be  6 \AA.
The initial error (9.8\%) is comparable to MACE without MP (8\%) but requires us to perform a hyperparameter search for the cutoff of the three-body terms. As shown in Fig. \ref{fig:rcut}, the error decreases as the cutoff of the two-body descriptors increases, reaching a value slightly below 5~\% when the cutoff is sufficiently large. In summary, naively increasing the two-body cutoff does not quite allow us to recover the 
accuracy of the long-range models. Nevertheless, the present observations have prompted us
to routinely use relatively larger cutoffs for the two-body descriptors (8~\AA), and
fairly small radial cutoffs for the three-body descriptors (5~\AA) in VASP as default.\\

Furthermore, irrelevant information worsens the condition number of the design matrix and the numerical stability of the problem.\\
Especially when considering long-range descriptors, it is crucial to carefully select the information to be included. Ideally, all relevant hyperparameters should be optimized by a grid search.

\subsection{Solid Sodium Chloride}

As a less demanding test, we investigate solid NaCl.  To construct a training set, a 128 atom fcc supercell was heated from 200 K to 900 K in 100,000 steps (1.5 ns) for an NPT ensemble.  The same hyperparameters as for liquid sodium chloride were used to fit the data, except that the forces and energies were weighted equally.

Table \ref{tab:res_nacl_solid} shows that the errors for forces and energies are about a factor of three smaller than for liquid NaCl, a finding we typically find: liquids are much more difficult to fit than solids. Again, the single-layer MACE model and the local model show rather similar errors, and the inclusion of long-range density descriptors reduces the error by almost a factor of 3. In this case,
LODE and the density descriptors show remarkably similar errors. As for liquid NaCl, increasing the cut-off radius for LODE and the number of basis functions worsens the results.

The performance of MACE with invariant and equivariant message passing is even better than for the liquid, improving by a factor of 2 compared to both LODE and the long-range density descriptors. An error well below 1~\% is indeed quite remarkable.
\\
\begin{table}[h]
\caption{\label{tab:res_nacl_solid}%
The root-mean-square percentage error (RMSPE) was calculated for the validation data of solid NaCl using combined density and radial LODE descriptors describing the electrostatic potential and a more flexible one as well as local-only descriptors and MACE \cite{MACE} with and without Message Passing (MP) as well as using invariant (inv.) and equivariant (equiv.) features. Radial cutoffs of typically 6~\AA $ $ are used. The results show that LODE and the density-based approach outperform the purely local method for real materials and including long-range interactions is essential.\\}

\begin{ruledtabular}
\begin{tabular}{lcc}
 &RMSPE E (\%)& RMSPE F (\%)\\
 \hline
 Local&0.7&4.2\\
 MACE no MP&0.8&3.8\\
 LODE minimal&0.4&1.4\\
 LODE flexible&0.6&2.6\\
 Density&0.4&1.5\\
 MACE inv.&0.3&0.7\\
 MACE equiv.&0.3&0.7\\
 \end{tabular}
\end{ruledtabular}
\end{table}

\subsection{Zirconia}\label{ZrO2}

Bulk zirconia ($\text{ZrO}_\text{2}$) is a material with a significant difference in electronegativity between its two atom types, but also exhibits strong covalent bonding as opposed to NaCl. It hence poses a more significant challenge than a simple ionic material. \\
The data set for this material consists of 592 distinct structures, each containing 32 Zr and 64 O atoms. The configurations are taken from VASP MD simulations. During these simulations, the zirconia was heated from 300K to 2800K using a Langevin thermostat in an NPT ensemble. For further details on the data set we refer to Ref.  \onlinecite{carlaZrO2}. \\
Again $20\%$ out of 592 training structures are used for validation.\\
The hyperparameters used for the short-range parts and MACE were identical to those employed for NaCl. However, for the density-based descriptor, choosing $N_{\text{max}} = 5$, $s = 1.5$, $\zeta_{\text{max}} = 2$ yields the best results.\\
For the "minimal" LODE approach, again $R_{\text{cut}} = 1~$\AA\, $N_{\text{max}} = 1$ and $\sigma = 0.3$
was used. Increasing the cutoff radius further to $R_{\text{cut}} = 5~$\AA\ and setting $N_{\text{max}} = 5$ (LODE flexible) does not significantly improve the results.
In this specific case, we also attempted to include three body descriptors with $L_{\text{max}} = 1$  (LODE three-body) as anisotropic dipole interactions may
play an important role (see below), but this also did not improve
the results significantly, as demonstrated in the Table \ref{tab:res_zro2}.
Here for the kernel methods the short- and long-range parts of the descriptors are weighted equally, which a hyper-parameter search showed to be optimal.\\

A comparison of the results for zirconia in Tab. \ref{tab:res_zro2} shows that MACE with MP yields the smallest error. However, even the purely local methods result in errors in the single-digit percentage range, indicating that the system is already fairly well-described using only short-range descriptors. The addition of long-range entries from the density-based and LODE method does not significantly improve the results.\\ 
\begin{table}[h]
\caption{\label{tab:res_zro2}%
The root-mean-square percentage error (RMSPE) was calculated for the validation data of ZrO$_2$ using combined density and LODE descriptors describing the electrostatic potential, a more flexible one and a three-body one, as well as local-only descriptors and MACE \cite{MACE}, with and without Message Passing (MP) as well as using invariant (inv.) and equivariant (equiv.) features. For a cutoff radius of $6$ \AA$ $ in the short-range part of the descriptor, there is hardly any difference between the three kernel methods and they do not reach the accuracy of MACE.\\}

\begin{ruledtabular}
\begin{tabular}{lcc}
 &RMSPE E (\%)& RMSPE F (\%)\\
\hline
Local&1.8&6.8\\
MACE no MP&1.7&6.0\\
LODE minimal&1.7&6.5\\
LODE flexible&1.8&6.2\\
LODE three-body&1.9&6.3\\
Density&1.4&6.2\\
MACE inv.&1.0&3.3\\
MACE equiv.&0.7&2.4\\
\end{tabular}
\end{ruledtabular}
\end{table}

As both the LODE (with two-body descriptors) and the density-based model describe only monopole-monopole interactions and yield results that are as good as those of the local description, it can be concluded that no significant monopole-monopole interactions are present in zirconia.\\
When comparing MACE with and without MP it is evident that the MP model improves the results. We speculate that, despite the absence of simple monopole-monopole interactions in zirconia, higher-order long-range electrostatic interactions, such as dipole-dipole interactions, play an important role. Although our results are indirect, we suspect that MACE (and other message-passing networks) can describe dynamical long-range dipole-dipole interactions.  Specifically, MP networks can learn dynamic, displacement-induced and direction-dependent dipoles on one site in early layers (e.g. MACE 1st MP layer) and deduce the interaction energies with other sites in later layers. We cannot see how a simple density-based descriptor or LODE can potentially model such interactions, and our empirical evidence also points against it.

\section{Conclusion}

MLFFs have advanced significantly over the past years, with the equivariant message passing networks in particular greatly improving prediction quality for materials where some form of long-range physics is involved. The consensus seems to be that for solid-state materials, message-passing networks improve the accuracy typically by a factor of two to three compared to more traditional invariant perceptrons or standard kernel-based approaches.
 
The exact reason for this is not yet fully understood. The present work aims to take a rational approach to the problem and attempts to propose long-range descriptors that are capable of describing a specific type of long-range interaction. Our starting point is the long-distance equivariant (LODE) approach, which is designed to describe electrostatic interactions between charges \cite{LODE} but also static dipoles \cite{lode2, huguenin2023physics}. The important difference is that we aimed to stay within the standard scheme to express the environment of each atom in a suitable set of invariant descriptors. It turns out that this is possible and only requires one to adopt suitable slowly decaying projectors. 

We show that the present long-range descriptor can achieve almost the same accuracy as the LODE descriptors for point charge models. Matter of fact, the learning efficiency is worse than for LODE, since the machine learning model has to determine from the data which linear combination of descriptors describes the Coulomb $1/r$ behavior. However, the present approach is more flexible as it can, by construction, describe any interaction, e.g. even a distance-dependent screened Coulomb interaction. We demonstrate this for a real material, liquid NaCl, where the long-range interactions are dominated by electrostatic screened interactions between point charges. For this particular test case, the present density-based descriptors combined with the usual short-range descriptors appear to outperform LODE combined with short-range descriptors.  Nevertheless, the model is found to be only as accurate as MACE and cannot improve upon the flexible message-passing model. For solid NaCl, 
LODE and the present long-range density descriptors perform very similarly and MACE clearly outperforms them.

For the second test material, various phases of ZrO$_2$, we find no improvement using long-range descriptors, and the performance of MACE cannot be matched. This indicates that the present electrostatic models (long-ranged density approach as well as LODE) still lack important physics. The likely explanation is that in ZrO$_2$ the dynamical (Born effective) charges are strongly anisotropic, i.e. moving in the O-Zr bond direction or orthogonally to them, respectively, gives very different Born effective charges. Such a behavior cannot be approximated
using monopole-monopole interactions only, and hence 
our surrogate model is unable to describe this by construction (it rather assumes an isotropic, possibly distance dependent interaction). 

Clearly, more work is needed to fully understand what kind of physics needs to be included to improve the short-range models that have dominated research over the last decade. We believe that this rational approach remains relevant, even though data-based flexible message-passing networks are currently outperforming the rationally designed surrogate models.

\section{Acknowledgements}
This research was funded in whole by the Austrian Science Fund (FWF) 10.55776/F81. For open access purposes, the author has applied a CC BY public copyright license to any author-accepted manuscript version arising from this submission. The presented computational results have been largely obtained using the Vienna Scientific Cluster (VSC).

\section{Author Declarations}
\subsection{Conflict of Interest}
The authors have no conflicts to disclose.

\section{Data Availability}
The datasets used/produced in this work can be freely accessed at \url{https://doi.org/10.5281/zenodo.13985979}.

\bibliography{apssamp}

\end{document}